\documentclass [] {aa} 
\usepackage{epsfig, graphicx,aalongtable}
\usepackage{natbib}
\usepackage{times}

\bibpunct{(}{)}{;}{a}{}{,}

\begin{document}


\title{The HARPS survey for southern extra-solar planets\thanks{Based 
	       on observations collected at La Silla Observatory, 
               ESO, Chile, with the HARPS spectrograph,
               at the 3.6-m ESO telescope (programs 073.D-0578 and 072.C-0488).}}

\subtitle{II. A 14 Earth-masses exoplanet around $\mu$\,Arae}

\author{N.C.~Santos\inst{1,3} \and
        F.~Bouchy\inst{2} \and
	M.~Mayor\inst{3} \and
	F.~Pepe\inst{3} \and
	D.~Queloz\inst{3} \and
	S.~Udry\inst{3} \and
	C.~Lovis\inst{3} \and
	M.~Bazot\inst{4} \and
	W.~Benz\inst{5} \and
	J.-L.~Bertaux\inst{6} \and
	G.~Lo Curto\inst{7} \and
	X.~Delfosse\inst{8} \and
	C.~Mordasini\inst{5} \and
	D.~Naef\inst{7,3} \and
	J.-P.~Sivan\inst{2} \and
	S.~Vauclair\inst{4}
	}

\offprints{Nuno C. Santos, \email{Nuno.Santos@oal.ul.pt}}

\institute{
        Centro de Astronomia e Astrof{\'\i}sica da Universidade de Lisboa,
        Observat\'orio Astron\'omico de Lisboa, Tapada da Ajuda, 1349-018
        Lisboa, Portugal
     \and
	Laboratoire d'Astrophysique de Marseille, Traverse du Siphon, 13013 Marseille, France
     \and
	Observatoire de Gen\`eve, 51 ch.  des 
	Maillettes, CH--1290 Sauverny, Switzerland
     \and
        Laboratoire d'Astrophysique, Observatoire Midi-Pyr\'en\'ees, 14 Avenue Edouard 
	Belin, 31400 Toulouse, France	
     \and
        Physikalisches Institut, Universit\"at Bern, Sidlerstrasse 5, CH-3012 Bern, Switzerland	
     \and
        Service d'A\'eronomie du CNRS, BP 3, 91371 Verri\`eres-le-Buisson, France	
     \and
        European Southern Observatory, Casilla 19001, Santiago 19, Chile 
     \and	
	Laboratoire d'Astrophysique de l'Observatoire de Grenoble, 414 rue de la piscine, 
	38400 Saint Martin d'H\`ere, France	
        }
	
\date{Received / Accepted } 

\titlerunning{The first 14 earth-mass exoplanet} 


\abstract{
In this letter we present the discovery of a very light planetary companion
to the star $\mu$\,Ara (HD\,160691). The planet orbits its host once every 9.5\,days,
and induces a sinusoidal radial velocity signal with a semi-amplitude 
of 4.1\,m\,s$^{-1}$, the smallest Doppler amplitude detected so far. These values 
imply a mass of $m_2\sin{i}$=14\,M$_{\oplus}$ (earth-masses). 
This detection represents the discovery of a planet with a mass slightly smaller 
than that of Uranus, the smallest ``ice giant" in our Solar System. Whether this
planet can be considered an ice giant or a super-earth planet is
discussed in the context of the core-accretion and migration models.
}

\maketitle

\section{Introduction}

The discovery of giant planets around other solar-type stars has opened
the way to a new era of planetary research. The new worlds present
a wide variety of orbital characteristics and minimum masses, and
9 years after the first announcement \citep[][]{May95}, some
of their properties are still defying the theories of planetary
formation. The increasing number of known systems is, however, giving the possibility
to explore their properties from a statistical point of 
view \citep[e.g.][]{San01,Zuc02,Udr03,Egg04}, and the observational and
theoretical approaches are now starting to converge \citep[e.g.][]{Tri02,Ali04,Ida04a}.

Recently, with the installation of the new HARPS spectrograph \citep[][]{Pep02} at 
the 3.6-m ESO telescope (La Silla, Chile) a significant quantitative advance has been possible.
This state of the art instrument is capable of attaining a precision better
than 1\,m\,s$^{-1}$. After only a few weeks of operation, it has discovered a first 
``hot-jupiter'' \citep[][]{Pep04} orbiting the K dwarf \object{HD\,330075}.
The level of precision in radial-velocity measurements achieved with HARPS
gives now, for the first time, the possibility of lowering significantly
the detection limit to the ``few-earth-mass" regime, provided that the signal induced 
by stellar oscillations can be reduced with the use of an appropriate 
observing strategy (Bouchy et al., in prep.).

In this letter we present the discovery of a $\sim$14-M$_{\oplus}$ short 
period (P$\sim$9.5\,days) extra-solar planet orbiting the star \object{$\mu$\,Ara},
a star that was already known to be orbited by a longer period giant planet \citep[][]{But01}. Together with the very low mass companion to 
\object{55\,Cnc} \citep[][]{McA04}, these are the only two sub-neptunian planets discovered 
to date. They are suspected to be earth-like rocky planets, orbiting solar-type stars.

\section{Stellar characteristics of \object{$\mu$\,Ara}}
\label{sec:star}

\object{$\mu$\,Ara} (\object{HD\,160691}, \object{HR\,6585}, \object{GJ\,691}) 
is a nearby V=5.12 magnitude southern G5V star in the constellation Ara, the Altar, 
and according to the Hipparcos catalog \citep[][]{ESA97}, it has a parallax of 65.5$\pm$0.8\,mas, 
which implies a distance from the Sun of 15.3\,pc, and an absolute magnitude
of M$_{\mathrm{v}}$=4.20. Its color index $B-V$ is 0.694.

From a HARPS spectrum with a S/N ratio of the order of $\sim$1000 (average of 275 
individual spectra), we have derived the stellar parameters for \object{$\mu$\,Ara} using a 
fully spectroscopic analysis \citep[][]{San04}. The resulting parameters 
(T$_{\mathrm{eff}}$, $\log{g}$, V$_{\mathrm{t}}$, [Fe/H])=(5813$\pm$40\,K, 4.25$\pm$0.07\,dex, 1.30$\pm$0.05\,km\,s$^{-1}$, $+$0.32$\pm$0.05\,dex), are in almost perfect agreement with the values published in \citet[][]{San04}, \citet[][]{Ben03}, and \citet[][]{Law03}. 
The surface gravity derived using the Hipparcos parallax and an effective temperature
of 5800\,K is 4.25\,dex \citep[see e.g.][]{San04}. 

\begin{figure}[t]
\psfig{width=\hsize,file=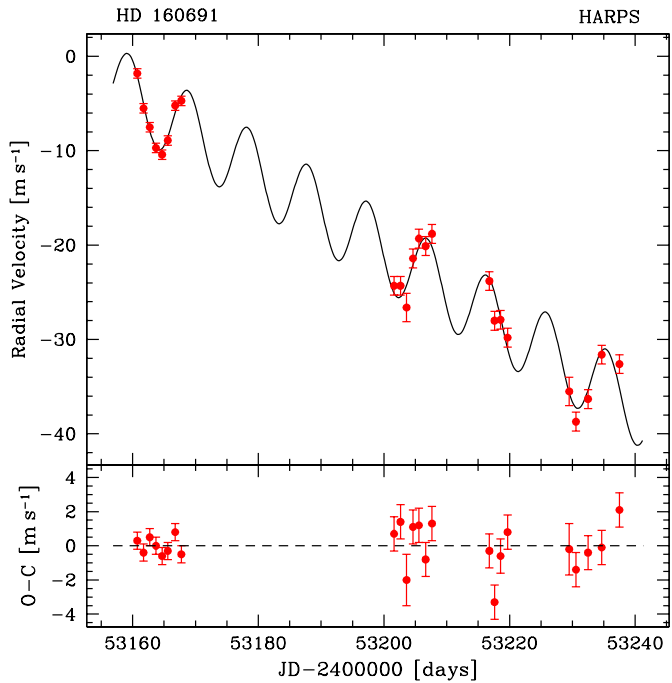}
\caption[]{HARPS radial-velocity measurements of \object{$\mu$\,Ara} as a function
of time. The filled line represents the best fit to the data, obtained
with the sum of a keplerian function and a linear trend, representing the effect 
of the long period companions to the system. The residuals of the fit, with 
an rms of only 0.9\,m\,s$^{-1}$, are shown in the lower panel.}
\label{fig:rv}
\end{figure}

Using the temperature, [Fe/H], absolute magnitude and bolometric correction 
\citep[][]{Flo96}, we derived a stellar mass of 1.10$\pm$0.05\,M$_{\sun}$ for \object{$\mu$\,Ara},
from an interpolation of the theoretical isochrones of \citet[][]{Sch93}. This is in excellent
agreement with the 1.08 and 1.14\,M$_{\sun}$ derived by \citet[][]{But01} and
\citet[][]{Law03}, respectively. Preliminary results from the asteroseismology 
analysis are also in excellent agreement with these values (Bazot et al., in prep.).

From the width of the CORALIE Cross-Correlation Function (CCF) we have computed a projected 
rotational velocity of 2.4\,km\,s$^{-1}$ for \object{$\mu$\,Ara} \citep[][]{San02b}. 
This value is in agreement with the low chromospheric activity level of the star, $\log{R'_{HK}}$=$-$5.034$\pm$0.006, obtained from the HARPS spectra. 
Similar values of $-$5.02 were obtained both from the CORALIE data \citep[][]{San00} 
and by \citet[][]{Hen96} at different epochs. The inactivity of this star is 
further supported by its low (and non-variable) X-ray luminosity \citep[][]{Mar02}, 
as well as by the lack of significant photometric variation in the Hipparcos data \citep[][]{ESA97}.

From the observed value of $\log{R'_\mathrm{HK}}$ we can 
infer an age above $\sim$2\,Gyr \citep[][]{Pac04} and a rotational period 
of $\sim$31\,days \citep[][]{Noy84}. This age is compatible with the 4.5\,Gyr 
obtained from an interpolation of theoretical isochrones \citep[e.g.][]{Law03}, and 
with the upper value for the lithium abundance $\log{\epsilon}(Li)$$<$0.86\,dex 
derived by \citet[][]{Isr04} for this dwarf. 

\begin{figure}[t]
\psfig{width=8.5cm,file=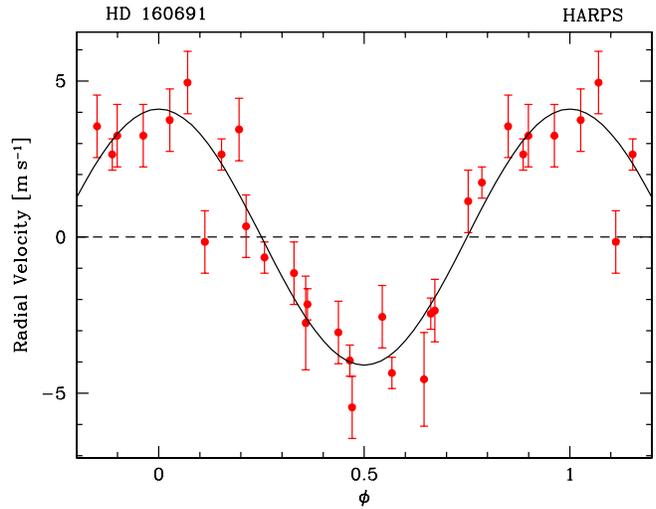}
\caption[]{Phase-folded radial-velocity measurements of \object{$\mu$\,Ara} after subtraction of
the linear trend shown in the upper panel of Fig.\,\ref{fig:rv}. In both panels the 
error bars represent the rms around the weighted average of the individual measurements 
for a given night.}
\label{fig:phas}
\end{figure}

\section{Radial velocities}
\label{sec:rv}

In June 2004, \object{$\mu$\,Ara} was intensively measured over 8 consecutive nights 
with the HARPS spectrograph as part of an asteroseismology program (Bouchy et al., in prep).
During each night, we obtained more than 250 spectra 
of this star, from which we derived accurate radial velocities. 
The average radial velocity for each night was then computed from a weighted average
of each individual value, its precision being limited by the uncertainty in the wavelength 
calibration\footnote{The nightly average of the HARPS radial velocities will 
be available in electronic form at CDS}.

The main motivation of this program was to study the possibility that the high 
metal content of the planet-host stars \citep[e.g.][and references therein]{Gon98,San01,San04} 
is due to the engulfment of metal rich planetary material into their convective envelopes.
Although current studies seem to favor that the observed ``excess" metallicity
reflects a higher metal content of the cloud of gas and dust that
gave origin to the star and planetary system, recent results have suggested
that this matter may still be unsettled \citep[e.g.][]{Vau04}.
The asteroseismology technique provides us with a good tool to possibly solve this
problem. As shown by \citet[][]{Baz04},
precise stellar oscillation measurements may be able to determine if there is
some metallicity gradient in the stellar interior, that could be a hint of strong
stellar ``pollution" events. The results of the asteroseismology campaign will
be presented in Bouchy et al. (in prep.) and Bazot et al. (in prep).

A first analysis of the data revealed what could be a periodic variation
with an amplitude of about 4\,m\,s$^{-1}$ (see Figs.\,\ref{fig:rv} and \ref{fig:phas}). 
As part of the HARPS GTO program, this star was then closely followed from July 14th to 
August 19th 2004 (16 radial-velocity measurements were obtained). 
Each night the radial velocity was measured from the average of about 
15 consecutive independent radial velocity estimates (computed from different spectra) 
taken during a period of $\sim$20\,minutes. 
This methodology makes it possible to average the radial-velocity variations 
due to stellar oscillations \citep[][]{May03b} -- see also Bouchy et al. (in prep.). 
As seen in Fig.\,\ref{fig:rv}, the measurements done during 
the first 8 nights (when the star was followed during the whole night) have a considerable 
lower rms around the best keplerian fit than the following measurements. This scatter 
results from the photon noise error ($\sim$20\,cm\,s$^{-1}$), the calibration uncertainty ($\sim$40\,cm\,s$^{-1}$), and from the stellar noise ($\sim$80\,cm\,s$^{-1}$) 
that is not completely averaged on the nights with only 15 radial velocity measurements 
(Bouchy et al., in prep.).


\object{$\mu$\,Ara} was previously announced to harbor a giant planet
in a long period ($\sim$740\,days) orbit \citep[][]{But01}.
This orbital solution has since been updated by \citet[][]{Jon02},
who found that the residuals of the radial-velocity planetary fit followed
a long term trend, due to the presence of a second body in the
system. 

\begin{table}[t]
\caption[]{
Orbital elements of the fitted 9.5-days period orbit and main planetary properties. }
\label{tab:planet}
\begin{tabular}{lr@{\,$\pm$\,}ll}
\hline
\noalign{\smallskip}
$P$                & 9.55                       & 0.03  & [d] \\
$T$                & 2453168.94                 & 0.05  & [d] \\
$e$        & 0.00  & 0.02         &     \\
$\omega$   & 4 & 2           & [deg] \\ 
$K_1$              & 4.1                        & 0.2   & [m\,s$^{-1}$] \\
$a_1\,\sin i$      & \multicolumn{2}{c}{0.5396} & [Gm]\\
$f_1(m)$           & \multicolumn{2}{c}{0.6869}         & [$10^{-13}\,M_{\odot}$]\\ 
$\sigma(O-C)$      & \multicolumn{2}{c}{0.9}            & [m\,s$^{-1}$]  \\    
$N$                & \multicolumn{2}{c}{24}             &   \\
$m_2\,\sin i$      & \multicolumn{2}{c}{14}             & [M$_\mathrm{\oplus}$]\\
$a$                & \multicolumn{2}{c}{0.09}           & [AU]\\
$T_{\mathrm{eq}}$ & \multicolumn{2}{c}{$\sim$900$^\star$}          & [K]\\
\noalign{\smallskip}
\hline
\end{tabular}
\\$^\star$ Equilibrium temperature computed with an albedo of 0.35.
\end{table}

In Fig.\,\ref{fig:long} we plot the radial-velocity measurements of
\object{$\mu$\,Ara} obtained during the last 6 years using three different instruments 
(see figure caption), as well as the best 2-keplerian fit. The orbit of the $\sim$740-day period planet 
(actually with a period of $\sim$660\,days) is confirmed. However, the orbital parameters of 
the second (longer period) companion are not well constrained; we find a strong degeneracy
between the derived orbital period and the value of the orbital excentricity, making it
possible to fit the data with the former parameter varying between $\sim$3000 and 
10000 days. Although not precisely determined, the mass of this companion remains
probably in the planet regime. Despite of the still unconstrained long period of this
outer companion, some stability studies of the system has been discussed 
\citep[e.g.][]{Goz03}.

\section{A 9.5-days period planet with 14\,Earth-masses}
\label{sec:planet}

\begin{figure}[t]
\psfig{width=\hsize,file=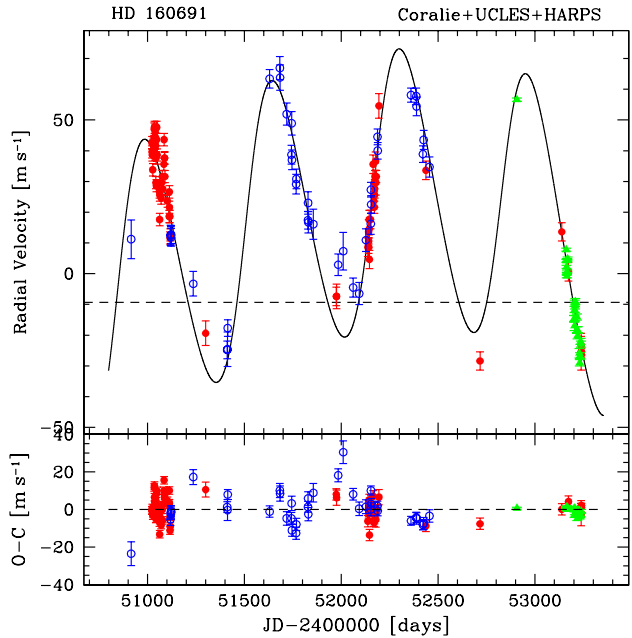}
\caption[]{Radial velocity measurements of \object{$\mu$\,Ara} 
obtained during the past 6 years with the CORALIE (dots) and HARPS spectrographs 
(open triangles), and by \citet[][]{Jon02} (open circles). The curve represents 
the best 2-body keplerian fit to the data. 
In the lower panel we present the rms around the fit. For the longer period
keplerian fit, the eccentricity was fixed to a value of 0.2.}
\label{fig:long}
\end{figure}

In Figs.\,\ref{fig:rv} and \ref{fig:phas} we present the HARPS radial-velocity measurements 
of \object{$\mu$\,Ara} as a function of time. In this figure, the curve represents 
the best fit to the data, obtained with the sum of a keplerian function and a 
linear trend. The derived slope of this trend is in agreement with the expected 
effect due to the longer period companions (see Fig.\,\ref{fig:long}).

The analysis of the radial velocity measurements reveals a variation 
with a period of 9.5 days, and a semi-amplitude of about 4\,m\,s$^{-1}$.
These values can be explained by the presence of a m$_2$\,$\sin{i}$ = 14\,M$_{\oplus}$ 
planet orbiting \object{$\mu$\,Ara} in a circular orbit. 

The residuals around the best fit to the HARPS data are flat, with a rms of only 
of 0.9\,m\,s$^{-1}$. This rms decreases to the calibration level (0.43\,m\,s$^{-1}$) 
for the first 8 nights, attesting the incredible precision of this instrument. 
Despite the low amplitude of the radial velocity signal, the false alarm probability that
it is due to random noise is lower than 1\%, as derived through a Monte-Carlo simulation.

From the stellar luminosity and effective temperature we can derive a
radius of $\sim$1.32 solar radii for \object{$\mu$\,Ara}. Combined with the rotational 
period of 31\,days (see Sect.\,\ref{sec:star}), this implies a rotational velocity of the order 
of 2.2\,km\,s$^{-1}$ for \object{$\mu$\,Ara}, close to the measured value 
$v\,\sin{i}$=2.4\,km\,s$^{-1}$. Supposing that the 
orbital plane is perpendicular to the stellar rotation axis, this means that
the orbital inclination $\sin{i}$ is close to unity, and that the
observed minimum mass for the planet is not very different from its real mass.

Using the HARPS spectra we have derived both an activity index, based on \ion{Ca}{ii}
H and K lines, and the bisector of the cross-correlation function from the individual
spectra. No correlation is found between these quantities and the radial velocities 
within the measurement precision.
Given the very low activity level of \object{$\mu$\,Ara} and the inferred 
rotational period of $\sim$30\,days, it is very unlikely that rotational 
modulation is capable of producing the observed stable periodic radial-velocity 
variation. Furthermore, to have a rotational period of 9.5\,days, this star 
would have to rotate at about 7\,km\,s$^{-1}$. Such a rotational velocity 
would imply a much younger age for \object{$\mu$\,Ara}, not compatible with 
its low level of activity.

The presence of a 14\,M$_{\oplus}$ planet around \object{$\mu$\,Ara} thus
remains the only credible explanation for the observed 9.5-days period radial-velocity
variation.


\section{Discussion}

As current planetary formation models are still far from being able to
account for all the amazing diversity observed amongst the exoplanets
discovered thus far, we can only speculate on the true nature of the
present object. 

First, given its location and the characteristics of the central star,
it is unlikely that this object was in fact a much more massive giant
planet which has lost a large fraction of its envelope over its 
lifetime. This is supported by the fact that more massive planets exist 
orbiting much closer to stars with similar characteristics and by 
calculations by \citet[][]{Bar04} and \citet[][]{Lec04} which show 
that only planets significanly less massive than Jupiter would evaporate 
at 0.09\,AU. Except if outward migration has occurred, we conclude 
that the mass of this object has always remained small.

To understand the consequences of this, it is necessary to recall that
in the current paradigm of giant planet formation, a core is formed
first through the accretion of solid planetesimals. Once this core
reaches a critical mass (m${_\mathrm{crit}}$), accretion of gas in a runaway fashion
becomes possible and the mass of the planet increases rapidly \citep[e.g.][]{Ida04b}. 
This therefore implies that the current object has never reached the critical
mass, for otherwise the planet would have become much more massive.
Furthermore, recent giant planet formation models including disk
evolution and migration \citep[][]{Ali04} have shown that these
effects greatly shorten the formation time. Hence, it is unlikely
that the planet has migrated over large distances before reaching its
present location. It was thus probably formed inside the ice 
radius \citep[$\sim$3.2\,AU --][]{Ida04a}, and its composition 
should be dominated by rocky (telluric) material. We note that the high [Fe/H] 
of \object{$\mu$\,Ara} makes this case possible \citep[][]{Ida04a}.
Curiously, with 14\,M$_{\mathrm{\oplus}}$ and a=0.09\,AU, this planet is near the 
borderline of the mass-period desert defined by \citet[][]{Ida04b}, where no 
planets are supposed to exist.

The above considerations lead us towards the following scenario 
for the formation of the present planetary system. The more 
massive planet, with the present $\sim$660 days period orbit, begins to form first 
and migrates inwards while growing in mass. Towards the end of the lifetime of the
disk, the smaller planet is formed inside the orbit of the larger one,
probably at a distance not exceeding 3\,AU. Thus, we expect this object
to have a massive, essentially rocky core (as opposed to icy), surrounded
by a gaseous envelope with $\sim$5-10\% of its mass. It therefore probably qualifies
as a super-Earth and not as a failed ice-giant.


The discovery of this extremely low-mass planet represents a new benchmark for
planet surveys, and demonstrates the ability of instruments
like HARPS to detect telluric planets with just a few times the mass of the Earth.
In the future these detections will give the possibility to study the low
end of the planetary-mass distribution. This kind of planets may be 
relatively common, as according to recent simulations \citep[][]{Ida04a}, very low-mass 
planets may be more frequent than the previously found giant worlds. 
This is further supported by the recent detection of a first neptunian planet 
in a short period orbit around 55\,Cnc \citep{McA04}\footnote{A
companion to the M-dwarf \object{GJ\,436} with a minimum mass $m_2\,\sin i$=21\,M$_{\oplus}$
was also announced by \citet[][]{But04} after the submission of the 
current letter.}. Such planets will be preferential targets for 
space missions like the photometric satellites COROT and Kepler. Furthermore, the discovery 
of such low mass planets around stars that have at least one more giant exoplanet, 
makes of these systems very interesting cases to understand the processes of planetary 
formation and evolution.

\begin{acknowledgements}
  We would like to thank Y. Alibert and S. Randich for the fruitful discussion. 
  We would like to thank the support from the Swiss National Science Foundation and
  the Portuguese Funda\c{c}\~ao para a Ci\^encia e a Tecnologia. S. Vauclair 
  acknowledges a grant from Institut Universitaire de France. This study benifited from the support of the HPRN-CT-2002-00308
European programme.
\end{acknowledgements}


\begin{thebibliography}{}

\bibitem[Alibert et al.(2004)]{Ali04}
Alibert, Y., Mordasini, C., Benz, W. 2004, A\&A, 417, L25

\bibitem[Baraffe et al.(2004)]{Bar04}
Baraffe, I., Selsis, F., Chabrier, G., et al. 2004, A\&A, 419, L13

\bibitem[Bazot \& Vauclair(2004)]{Baz04}
Bazot, M., \& Vauclair, S. 2004, A\&A, submitted (astro-ph/0407544)

\bibitem[Bensby et al.(2004)]{Ben03}
Bensby, T., Feltzing, S., \& Lundstr\"om, I. 2003, A\&A, 410, 527

\bibitem[Butler et al.(2004)]{But04}
Butler, R.P., Vogt, S., Marcy, G., et al. 2004, ApJ, in press

\bibitem[Butler et al.(2001)]{But01}
Butler, R.P., Tinney, C.G., Marcy, G., et al. 2001, ApJ, 555, 410


\bibitem[Eggenberger et al.(2004)]{Egg04}
Eggenberger, A., Udry, S., \& Mayor, M. 2004, A\&A, 417, 353

\bibitem[ESA(1997)]{ESA97} 
ESA 1997, The Hipparcos and Tycho Cat., ESA SP-1200


\bibitem[Flower(1996)]{Flo96} 
Flower, P.J. 1996, ApJ 469, 355

\bibitem[Gonzalez(1998)]{Gon98} 
Gonzalez, G. 1998, A\&A, 334, 221

\bibitem[Gozdziewski et al.(2003)]{Goz03} 
Gozdziewski, K., Konacki, M., \& Maciejewski, A. 2003, ApJ, 594, 1019

\bibitem[Henry et al.(1996)]{Hen96} 
Henry, T.J., Soderblom, D.R., Donahue, R.A., \& Baliunas, S.L. 1996, AJ, 111, 439

\bibitem[Ida \& Lin(2004a)]{Ida04a}
Ida, S., \& Lin, D.N.C. 2004a, ApJ, in press (astro-ph/0408019)

\bibitem[Ida \& Lin(2004b)]{Ida04b}
Ida, S., \& Lin, D.N.C. 2004b, ApJ, 604, 388

\bibitem[Israelian et al.(2004)]{Isr04}
Israelian, G., Santos, N.C., Mayor, M., \& Rebolo, R. 2004, A\&A, 414, 601

\bibitem[Jones et al.(2002)]{Jon02}
Jones, H.R.A., Butler, R.P., Marcy, G.W., et al. 2002, MNRAS 337, 1170

\bibitem[Laws et al.(2003)]{Law03}
Laws, C., Gonzalez, G., Walker, K., et al. 2003, AJ 125, 2664

\bibitem[Lecavelier des Etangs et al.(2004)]{Lec04}
Lecavelier des Etangs, A., Vidal-Madjar, A., McConnell, J.C., \& H\'ebrard, G. 2004, A\&A, 418, L1


\bibitem[Marino(2002)]{Mar02}
Marino, A., Micela, G., Peres, G., \& Sciortino, S. 2002, A\&A, 383, 210


\bibitem[Mayor et al.(2003)]{May03b}
Mayor, M., Pepe, F., Queloz, D., et al. 2003, The ESO Messenger, 114, 20

\bibitem[Mayor \& Queloz(1995)]{May95}
Mayor, M., \& Queloz, D. 1995, Nature, 378, 355

\bibitem[McArthur et al.(2004)]{McA04}
McArthur, B., Endl, M., Cochran, W., et al., ApJ, in press

\bibitem[Noyes et al.(1984)]{Noy84} 
Noyes, R.W., Hartmann, L.W., Baliunas, S.L., et al. 1984, ApJ, 279, 763

\bibitem[Pace \& Pasquini(2004)]{Pac04}
Pace, G., \& Pasquini, L. 2004, A\&A, in press (astro-ph/0406651)

\bibitem[Pepe et al.(2004)]{Pep04}
Pepe, F., Mayor, M., Queloz, D., et al. 2004, A\&A, 423, 385

\bibitem[Pepe et al.(2002)]{Pep02}
Pepe, F., Mayor, M., Rupprecht, G., et al. 2002, The ESO Messenger, 110, 9


\bibitem[Santos et al.(2004)]{San04}
Santos, N.C., Israelian, G., \& Mayor, M. 2004, A\&A, 415, 1153



\bibitem[Santos et al.(2002)]{San02b}
Santos, N. C., Mayor, M., Naef, D., et al. 2002, A\&A, 392, 215

\bibitem[Santos et al.(2001)]{San01}
Santos, N.C., Israelian, G., \& Mayor, M. 2001, A\&A, 373, 1019

\bibitem[Santos et al.(2000)]{San00}
Santos, N. C., Mayor, M., Naef, D., et al. 2000, A\&A, 361, 265

\bibitem[Schaerer et al.(1993)]{Sch93} 
Schaerer, D., Charbonnel, C., Meynet, G., Maeder, A., \& Schaller, G. 1993, A\&AS, 102, 339

\bibitem[Trilling et al.(2002)]{Tri02}
Trilling, D., Lunine, J., \& Benz, W. 2002, A\&A, 394, 241

\bibitem[Udry et al.(2003)]{Udr03}
Udry, S., Mayor, M., \& Santos, N.C. 2003, A\&A, 407, 369

\bibitem[Vauclair(2004)]{Vau04}
Vauclair, S. 2004, ApJ, 605, 874

\bibitem[Zucker \& Mazeh(2002)]{Zuc02}
Zucker, S., \& Mazeh, T. 2002, ApJ, 568, L113

\end{thebibliography}
\end{document}